\documentclass[showpacs,epsf,twocolumn]{revtex4-1}
\usepackage{float}
\usepackage{color,graphicx}
\usepackage{hyperref}
\begin{document}
\title{Tuning the scattering mechanism in three-dimensional Dirac semimetal Cd$_{3}$As$_{2}$}

\author{A. Pariari, N. Khan, R. Singha, B. Satpati and P. Mandal}

\affiliation{Saha Institute of Nuclear Physics, HBNI, 1/AF Bidhannagar, Kolkata 700 064, India}
\date{\today}

\begin{abstract}
To probe the charge scattering mechanism in Cd$_{3}$As$_{2}$ single crystal, we have analyzed the temperature and magnetic field dependence of the Seebeck coefficient ($S$). The large saturation value of $S$ at high field clearly demonstrates the linear energy dispersion of three-dimensional Dirac fermion. A wide tunability of the charge scattering mechanism  has been realized by varying the strength of the magnetic field and carrier density via In doping. With the increase in magnetic field, the scattering time crosses over from being nearly energy independent to a regime of linear dependence.  On the other hand, the scattering time enters into the inverse energy-dependent regime and the Fermi surface strongly modifies with  2\% In doping at Cd site. With  further increase in In content from 2 to 4\%, we did not observe any Shubnikov-de Haas  oscillation up to 9 T field, but the magnetoresistance is found to be quite large as in the case of undoped sample.\\
\end{abstract}
\maketitle
\section{introduction}
The discovery of Dirac and Majorana  fermionic excitations  has opened up a new avenue  of research on the relativistic particles in condensed-matter systems \cite{castro,been}. This discovery not only quenches our thirst for fundamental physics but also provides the possibility  of  technological applications \cite{novo,naya}. There are compelling experimental evidences for the three-dimensional (3D) Dirac semimetallic phase in Cd$_{3}$As$_{2}$ and Na$_3$Bi \cite{wang2,wang1,liu,xu,liu1,neu,boris,jeon}. Unlike topological insulators \cite{hasan,qi}, these Dirac semimetals exhibit  linear energy dispersion relation in the bulk similar to graphene \cite{castro}, whereas the surface-state is topology protected Fermi arc. After the discovery of Dirac fermionic excitations in Cd$_{3}$As$_{2}$, more attention has been given on the electronic transport properties for understanding the nature and origin of ultrahigh magnetoresistance and mobility of charge carriers, as well as the Fermi surface (FS) geometry \cite{he,cao,tian,nara,feng,Zhao,arnab}. It has been predicted theoretically \cite{wang1,wang2} that the breaking of time reversal symmetry by external magnetic field rearranges the Fermi surface of Cd$_3$As$_2$. From the magnetoresistivity and Hall measurements, a strong field dependence of scattering time  has been observed and this behavior has been ascribed to the field-induced changes in the Fermi surface \cite{tian}. In this context, we would like to focus on some important differences  between the FS  proposed from angle-resolved photoemission spectroscopy (ARPES) and quantum oscillation measurements. ARPES results show that FS of Cd$_3$As$_2$ consists of two ellipsoids with negligible anisotropy or almost spherical \cite{liu1,boris}. A 3D plot of spectra intensity clearly shows linear dispersion with little anisotropy along two perpendicular directions in [111] plane \cite{liu1}. Whereas the Shubnikov-de Haas (SdH) \cite{nara,rose,Zhao} and de Haas-van Alphen (dHvA) \cite{arnab,hide} oscillations reveal anisotropic FS with different frequencies, Fermi velocities and carrier effective mass.  Furthermore, no surface experiment has been able to detect the Lifshitz transition up to 300 meV  [\cite{boris,liu1}]. On the other hand, the Lifshitz transition and, as a consequence, the FS nesting has been detected both from the SdH oscillation \cite{Zhao,enza} and dHvA effect \cite{arnab} well below 300 meV, which are consistent with the theoretical predication. Band structure calculations show that this transition is expected to occur at around $\sim$133 meV \cite{feng}. As quantum oscillations are observed in presence of high magnetic field, the associated features from this kind of study may  arise as a result of reconstruction of the FS by magnetic field. So, it is worthwhile to study how this change in the Fermi surface is reflected in the scattering of charge carriers and which relaxation process dominates the charge transport at high magnetic field. Doping also affects the Fermi surface by either reducing or enlarging its area. Thus, it is important to study the effect of doping and magnetic field simultaneously on the charge transport mechanism. The measurement of resistivity alone is not sufficient to understand the details of the scattering mechanism. Thermoelectric power ($S$) has been used as a powerful tool to probe the relaxation process in metals and semiconductors because it provides complementary information to resistivity due to its proportionality with the energy-derivative of the electrical  conductivity.  Indeed, the inverse square-root dependence of the thermoelectric power on carrier density clearly reflects the linear dispersion relation in graphene which is the fingerprint of massless Dirac fermions \cite{wei}.\\

In this work, we present a thorough study on thermoelectric properties of Cd$_{3}$As$_{2}$ to probe the possible scattering mechanisms. We have shown that the relaxation process can be widely tuned upon carrier doping and by applying external magnetic field, which is a step forward  towards understanding the material.\\
\section{SAMPLE PREPARATION, CHARACTERIZATION AND EXPERIMENTAL DETAILS}
\begin{figure}
\includegraphics[width=0.4\textwidth]{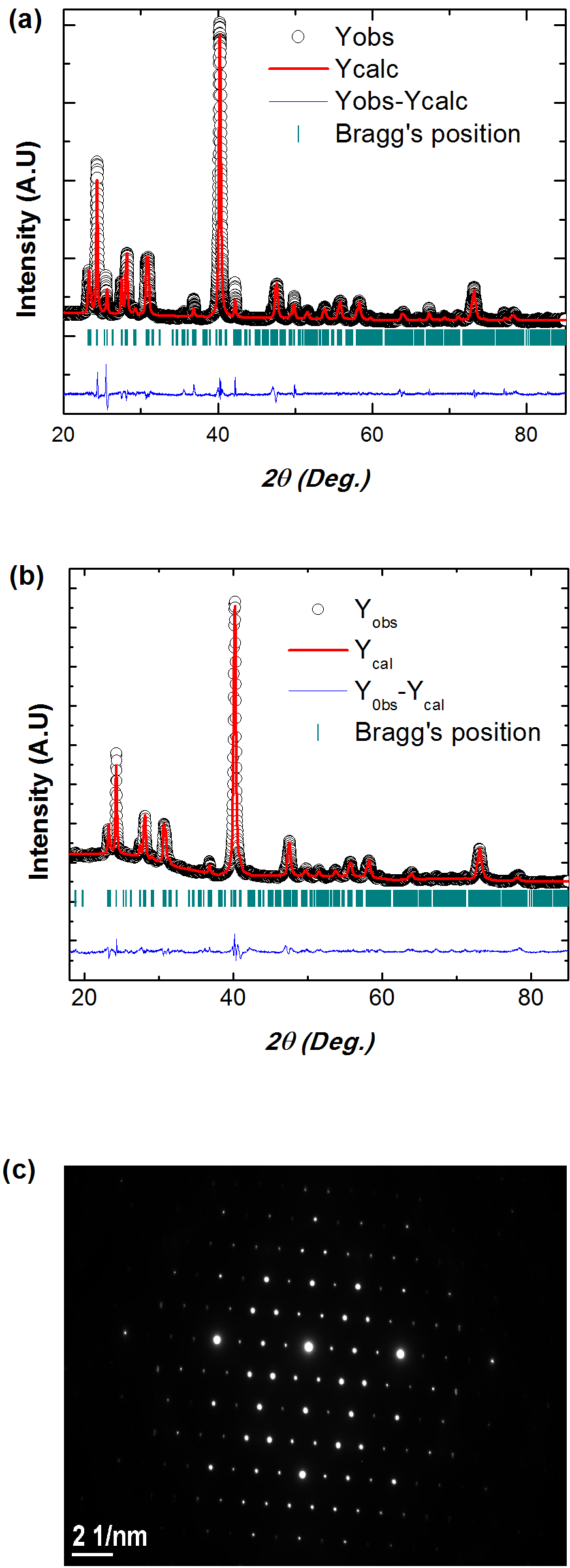}
\caption{(Color online) (a) X-ray diffraction pattern of powdered single crystals of (Cd$_{0.98}$In$_{0.02}$)$_3$As$_{2}$. Black (Y$_{obs}$), experimental data; red (Y$_{cal}$), the calculated pattern; blue (Y$_{obs}$-Y$_{cal}$), the difference between experimental and calculated intensities; green, the Bragg positions. (b) X-ray diffraction pattern of powdered single crystals of (Cd$_{0.96}$In$_{0.04}$)$_3$As$_{2}$. (c) Selective area electron diffraction (SAD) pattern obtained through HRTEM measurement for (Cd$_{0.98}$In$_{0.02}$)$_3$As$_{2}$. }\label{rh}
\end{figure}
Single crystals of Cd$_{3}$As$_{2}$ were synthesized by  chemical vapor transport technique. The details of sample preparation and characterization have been described in our earlier work \cite{arnab}. As the ionic  radii of cadmium and indium  are close to each other, we choose to dope In at Cd site in Cd$_{3}$As$_{2}$  to minimize the doping induced lattice disorder. Single crystals of (Cd$_{1-x}$In$_x$)$_3$As$_{2}$ with $x$=0.02 and 0.04 were prepared using the same technique as that for the undoped one. Phase purity and the structural analysis of the samples were done using the high-resolution powder x-ray diffraction (XRD) technique (Rigaku, TTRAX II) with Cu-K$_{\alpha}$ radiation. Figs. 1(a) and (b) show the room-temperature x-ray diffraction pattern for the powdered samples of single crystals with $x$=0.02 and 0.04, respectively.  Within the resolution of XRD, we have not observed any peak due to the impurity phase as a result of In doping. Using the Rietveld profile refinement program for the diffraction patterns, we have calculated the lattice parameters $a$$=$$b$$=$12.643 {\AA} and $c$$=$25.440 {\AA} for $x$=0.02 while $a$$=$$b$$=$12.663 {\AA} and $c$$=$25.466 {\AA} for $x$=0.04 with space-group symmetry $I_{41}/acd$. These values of lattice parameters are very close to that for  Cd$_{3}$As$_{2}$ \cite{arnab}.  Selective area electron diffraction (SAD) for the doped single crystal has been done using the high resolution transmission electron microscopy (HRTEM) in FEI, TECNAI G$^2$ F30, S-TWIN microscope operating at 300 kV equipped with a GATAN Orius SC1000B CCD camera.  Fig. 1(c) shows the SAD pattern for the $x$=0.02 crystal. Very clear periodic diffraction spots in the SAD pattern implies well-stacked crystal planes and absence of any crystal defects in the doped compound.  The  thermoelectric power (Seebeck coefficient) and magneto-transport measurements on undoped and doped Cd$_{3}$As$_{2}$ single crystals were done by four-probe technique using thermal and ac transport measurement options, respectively, in  physical property measurement system (Quantum Design). The typical dimensions of the undoped, 2\% In-doped and 4\% In-doped samples, used in both the thermoelectric power and transport measurements, are $\sim$3$\times$2$\times$0.55 mm$^3$, $\sim$3$\times$1.5$\times$0.6 mm$^3$ and $\sim$3.5$\times$1.5$\times$0.25 mm$^3$, respectively. All the measurements were carried out by applying magnetic field along [100] direction. Both the current and temperature gradient are along [012] direction; perpendicular to the applied magnetic field. Though several single crystals have been studied, we present the data for one single crystal as a representative for each composition. Qualitatively similar behavior has been observed for other crystals.\\

\section{RESULTS AND DISCUSSION}
\subsection{Resistivity and Hall resistance of  Cd$_{3}$As$_{2}$ single crystal}
\begin{figure}
\includegraphics[width=0.4\textwidth]{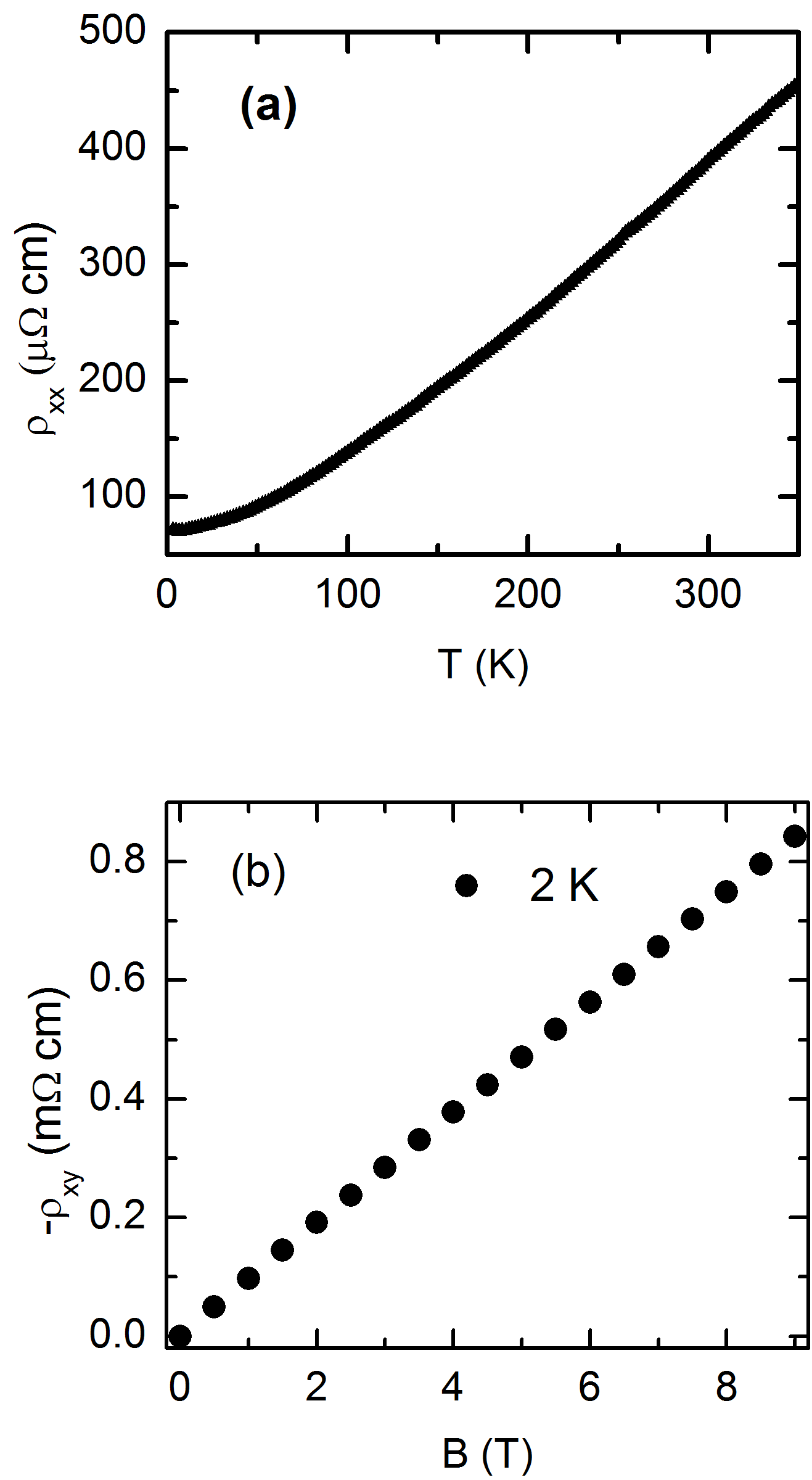}
\caption{(Color online) (a) Temperature dependence of  resistivity ($\rho$$_{xx}$) of Cd$_{3}$As$_{2}$ single crystal from 2 to 350 K.  (b) The field dependence of the Hall resistance at 2 K.}\label{rh}
\end{figure}
The temperature dependence of  resistivity ($\rho$$_{xx}$) for a Cd$_{3}$As$_{2}$ single crystal is shown in Fig. 2(a). Over the entire temperature range from 350 to 2 K, $\rho$$_{xx}$ exhibits weakly metallic behavior ($d\rho/dT$$>$0). The upward curvature of $\rho$$_{xx}$ versus $T$ curve suggests that $\rho$$_{xx}$ exhibits superlinear temperature dependence.  In Fig. 2(b),  the Hall resistivity ($\rho_{xy}$) is plotted as a function of magnetic field at 2 K. Figure shows that $\rho_{xy}$ is negative and increases linearly with field.  From the  slope of the Hall resistivity, the density of charge carrier  ($n$) is calculated to be $\sim$6.8$\times$10$^{18}$ cm$^{-3}$. Using the value of resistivity, we have calculated carrier mobility ($\mu$) at 2 K $\sim$1.3$\times$10$^4$ cm$^2$ V$^{-1}$ s$^{-1}$.  These values are close to the earlier reports \cite{he,cao,tian,nara}.\\
\subsection{Magnetoresistance and quantum oscillations in Cd$_{3}$As$_{2}$ single crystal}
\begin{figure}
\includegraphics[width=0.4\textwidth]{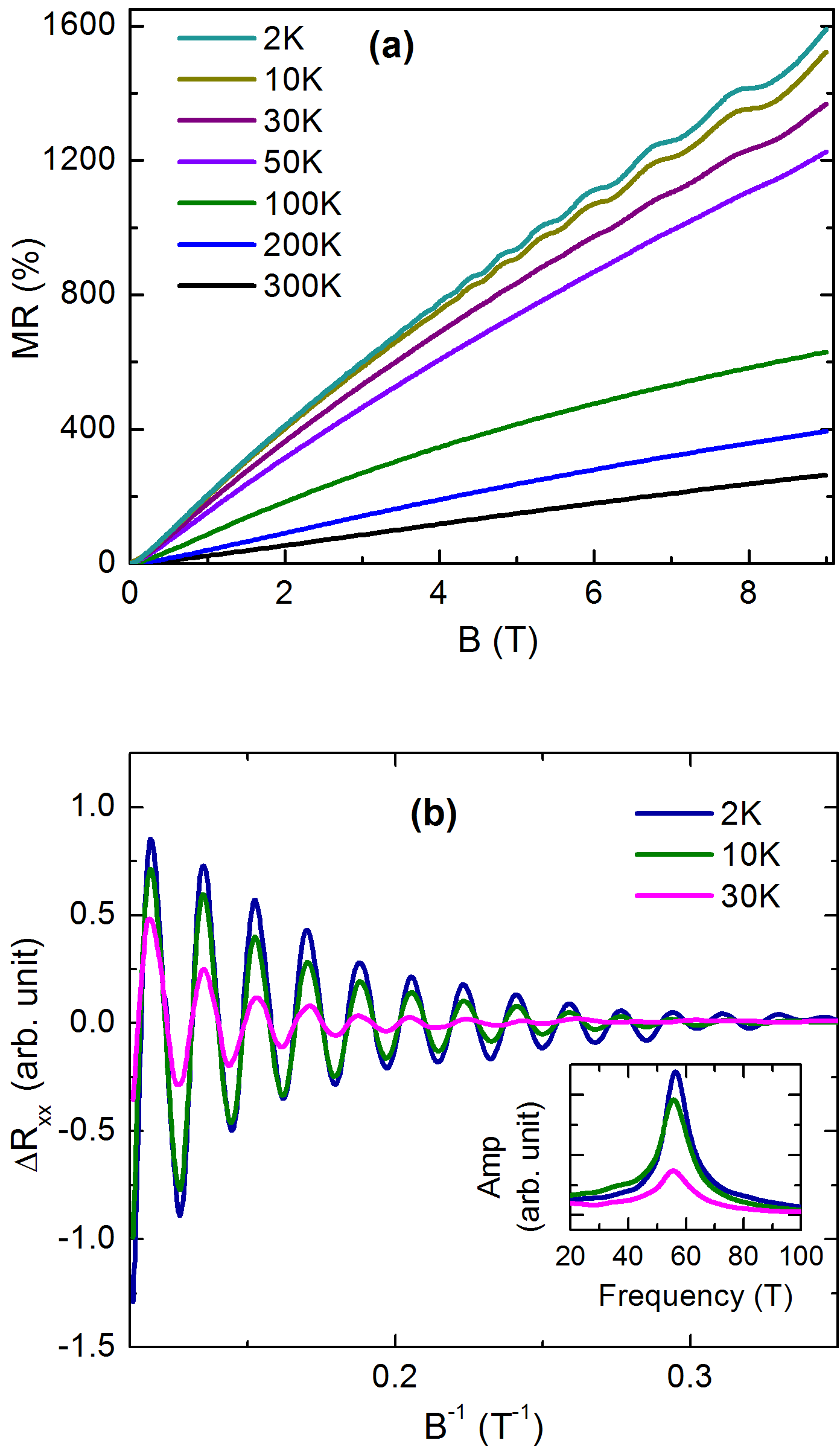}
\caption{(Color online) (a) Magnetoresistance, which is defined as [$\rho$$_{xx}$($B$) - $\rho$$_{xx}$(0)]/$\rho$$_{xx}$(0), of Cd$_{3}$As$_{2}$ single crystal at some representative temperatures from 2 to 300 K, when the field was applied  along the  [100] direction. (b) The oscillatory component $\Delta$$R_{xx}$ of MR in arbitrary unit (after subtracting a smooth background) as a function of  1/$B$; the inset shows the oscillation frequency after fast Fourier transform. }\label{rh}
\end{figure}
Figure 3(a) displays the  magnetic field ($B$) dependence of magnetoresistance (MR) of Cd$_{3}$As$_{2}$ crystal at several representative temperatures in the range 2-300 K, where MR is defined as [$\rho$$_{xx}$($B$) - $\rho$$_{xx}$(0)]/$\rho$$_{xx}$(0). Even at room temperature and 9 T magnetic field, MR is large ($\sim$260$\%$) and shows no sign of saturation. Except at low field, MR is approximately linear in $B$.  With decreasing temperature,  MR increases rapidly, and   reaches  $\sim$1600$\%$ at 2 K and 9 T. MR shows very clear Shubnikov-de Haas oscillation, traceable at a field as low as 3 T. $\Delta$$R_{xx}$, obtained after subtracting a smooth background from $R$($B$), is plotted in Fig. 3(b) as a function of 1/$B$. The fast Fourier transform spectra of $\Delta$$R_{xx}$ versus 1/$B$ curve shows a single oscillation frequency at around 56 T [inset of Fig. 3(b)]. The details of quantum oscillation analysis for the undoped sample have been reported in our earlier work \cite{arnab}.

\subsection{Temperature and magnetic field dependence of the Seebeck coefficient of Cd$_{3}$As$_{2}$}
\begin{figure}
\includegraphics[width=0.4\textwidth]{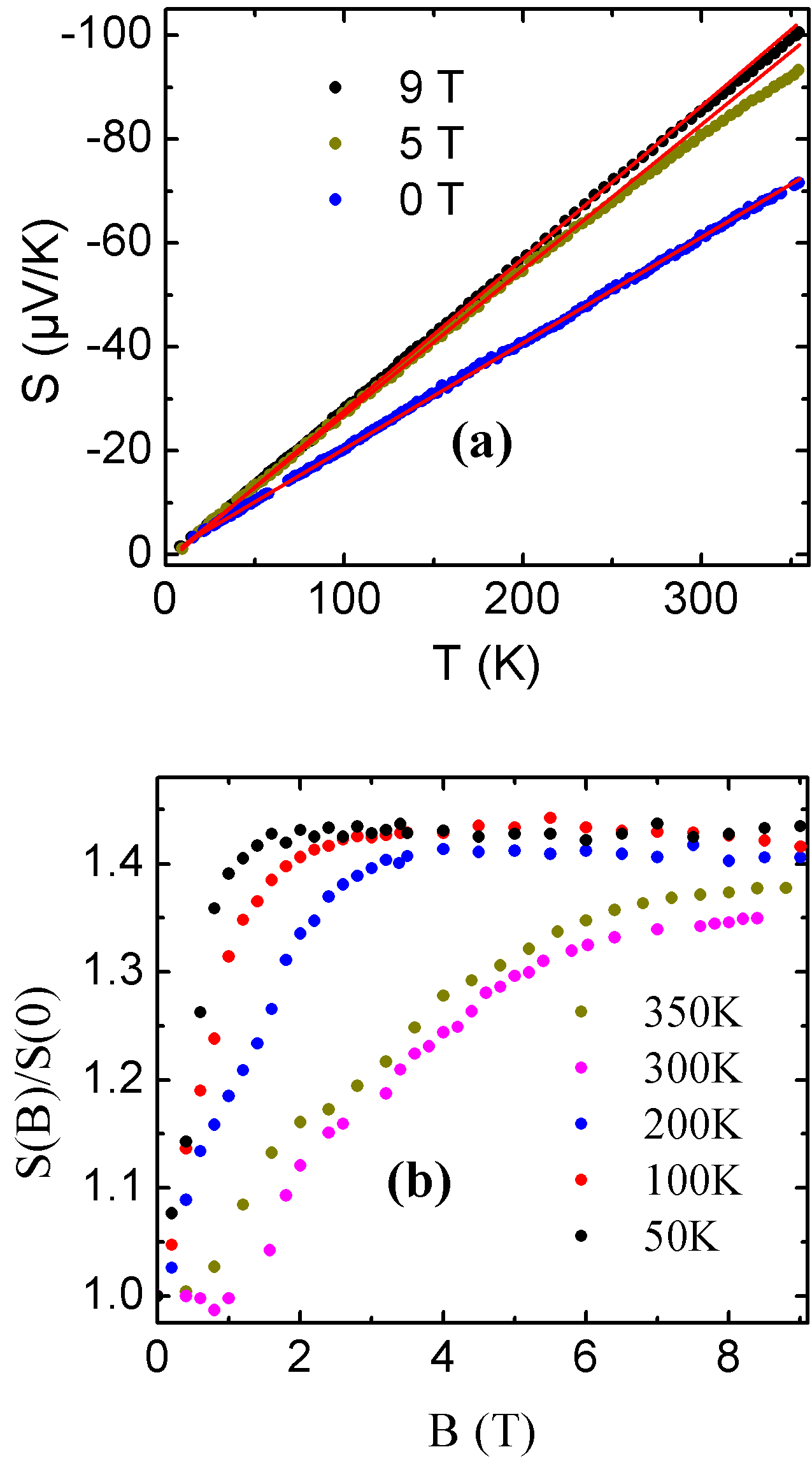}
\caption{(a) Temperature dependence of the Seebeck coefficient ($S$) of Cd$_{3}$As$_{2}$ single crystal up to 350 K at 0, 5  and 9 T fields. Solid line shows the linear $T$ dependence of $S$. (b) Magnetic field dependence of the normalized thermoelectric power at different temperatures between 50 and 350 K up to 9 T,  where $S$(0) is the zero-field Seebeck coefficient at the respective temperature.}\label{rh}
\end{figure}
In Fig. 4(a), the Seebeck coefficient  for Cd$_{3}$As$_{2}$ is plotted  as a function of $T$ up to 350 K both in presence and absence of external magnetic field.  The negative sign of $S$ implies  that the charge transport is dominated by electrons, which is consistent with the negative sign of the Hall coefficient. Remarkably, $S$ shows a  linear $T$ dependence almost up to 350 K at zero and 9 T magnetic field, whereas at 5 T, it shows a weak sublinear behavior at high temperature. A qualitatively similar temperature dependence of $S$ has been observed earlier in polycrystalline sample over the temperature range 77-270 K in absence of external magnetic field \cite {masu}. It is interesting to compare the temperature dependence of the Seebeck coefficient of Cd$_{3}$As$_{2}$  with that of graphene because of their striking similarities in electronic properties and band dispersion.  The linear behavior of the Seebeck coefficient  over a wide temperature range has also been reported for graphene \cite{yu,wei}. In graphene, however, both the type  and density of charge carrier  can be tuned by applying gate voltage. Figure 4(b) shows the magnetic field evolution of  $S$ at some representative temperatures. $S$ increases monotonically with field and tends to saturate at high  fields. At 350 K, about 30$\%$ increase in $S$ is observed at 5 T and, above 8 T, $S$ tends to saturate at $\sim$95 $\mu$V/K. For $T$$<$ 200 K, as the relative enhancement of $S$ from its zero-field value depends very weakly on temperature above $\sim$2 T, the slope of $S$($T$) curve in the low-temperature region converges faster with field to a definite value. The weak sublinear behaviour of $S$($T$) at 5 T in the high temperature region  [Fig. 4(a)] can be explained from such kind of field dependence of $S$. \\

Korenblit and Sherstobitov developed a semiclassical theory for the transport phenomenon for InSb-type degenerate semiconductors, assuming that the change in energy of the electrons on scattering is less than 2$k_BT$ \cite{koren}. In past, it was believed that Cd$_{3}$As$_{2}$ belongs to the same family as InSb and the validity of the above approach has been experimentally verified  for Cd$_{3}$As$_{2}$ \cite{gold2}. According to this theory, the saturation value of $S$ ($S$$_{\infty}$) at high field, in terms of the energy ($\varepsilon$) dependence of  the electron concentration ($p$) at the Fermi energy ($E_F$), is given by \cite{koren,gold2}
\begin{equation}
S_{\infty}=\frac{\pi^2k_B}{3e}\frac{k_BT}{E_F}\frac{d\ln p}{d\ln \varepsilon}.
\end{equation}
It was also assumed that $p \propto \varepsilon^s$, where $s$ determines the nature of energy dispersion of the system. Thus, the above expression simplifies to, $S_{\infty}$=$\frac{\pi^2k_B}{3e}\frac{k_BT}{E_F} s$. In three dimensions, $s$ is 3/2 for a usual parabolic band and 3 for a linear band. For Cd$_{3}$As$_{2}$ crystal, we have deduced $E_F$$\sim$270 meV from  the magnetotransport and magnetization measurements \cite{arnab}. This value of $E_F$  is close to that determined by several other groups  from magnetotransport studies \cite{he,cao,tian,chen}. Now, using the experimental value of $S_{\infty}$ at 350 K, we find $s$ $\sim$ 2.97. The field dependence of $S$ has also been analyzed for temperatures below 350 K. With decreasing $T$, $S$ tends to saturate at a lower field.  The value of $S$, however, turns out to be very close to 3 at all the temperatures.  Thus, the field dependence of the Seebeck coefficient clearly demonstrates the linear dispersion of 3D Dirac fermion in Cd$_{3}$As$_{2}$. In graphene, by tuning the carrier density through gate voltage, an inverse square-root dependence of $S$ on carrier density has been observed, which is a signature of the linear dispersion in this compound \cite{wei}.\\
\subsection{Analysis of the magnetic field evolution of the scattering time in Cd$_{3}$As$_{2}$}
For $T$$<<$$T_F$, where $T_F$ is the Fermi temperature, the linear $T$ dependence of $S$ has been ascribed to the Mott formula,
\begin{equation}
S=\frac{\pi^2k_B^2}{3e}\frac{T}{\sigma(\mu)}\frac{\partial\sigma(\varepsilon)}{\partial\varepsilon}|_{E_F},
\end{equation}
where $\sigma(\varepsilon)$ is the energy-dependent conductivity. $\sigma(\varepsilon)$  can be expressed in terms of Fermi velocity ($v_F$), density of states ($D$) and energy-dependent scattering time $\tau$; $\sigma(\varepsilon)=e^2v_F^2D(\varepsilon)\tau(\varepsilon)/2$ \cite{hwang}. For graphene, $D(\varepsilon)$=$g|\varepsilon|/2\pi\hbar^2v_F^2$ where $g$ is the total degeneracy.  Assuming the energy-dependent scattering time $\tau \propto \varepsilon^m$, one gets $S=\frac{\pi^2k_B}{3e}\frac{k_BT}{E_F}(m+1)$.\\

It has been shown that the Mott formula for graphene holds good up to temperature $T\sim$ 0.2$T_F$  \cite{yu,wei}. As $D(\varepsilon)=\frac{\varepsilon^2}{2\pi^2\hbar^3v_F^3}$ for a 3D Dirac system, considering Eq. (2) and the energy-dependent scattering time $\tau \propto \varepsilon^m$, we get
\begin{equation}
S=\frac{\pi^2k_B}{3e}\frac{k_BT}{E_F}(m+2).
\end{equation}
We have analyzed the linear $T$ dependence of $S$ using Eq. (3) and deduced  $m$$\sim$0.15 at zero field, i.e, $\tau$ is very weakly energy dependent. Similarly, the values of $m$, determined from the linear region of $S$($T$) curves at 5  and 9 T, are found to be very close to 1. The energy independence of $\tau$ at zero field is consistent with the random mass model of Dirac fermion, where a randomly fluctuating gap is introduced by randomly distributed scatterers (i.e. disorder) \cite{zeig,pall}. Though the linear $T$ dependence  of $S$ has been  predicted for the charged impurity ($m$=2) and short-range disorder ($m$=-2) scattering,  we observe a quite different energy dependence of $\tau$ for Cd$_3$As$_2$ \cite{rex}. However, in graphene, the dominant transport mechanism is the screened Coulomb scattering from charged impurities \cite{tan}. In Cd$_3$As$_2$, we observe that $m$ increases with field and tends to saturate at $m$=1 at high fields.\\

The magnetic field dependence of $m$ can be understood qualitatively from  Fig. 4(b). In the high-field region, where $S$ shows a saturation-like behavior, the value of $m$, determined from Eq. 3, is very close to 1. However, the field above which $S$ starts to saturate is extremely sensitive to temperature.  At low temperature, the saturation occurs at a relatively small applied field. For example, at 50 K,  $S$ increases very rapidly with the application of field and becomes almost independent of $B$ above 2 T, i.e., $m$ increases sharply from 0.15 at zero field to about 1 at 2 T. As $S$($B$) curve shifts progressively leftward with decreasing $T$, $m$ is expected to increase very sharply and reach 1 at a much smaller field strength when the temperature is decreased further below 50 K. This implies that the relaxation process at high fields is dominated by the  unscreened charged impurity \cite{hwang}. The evolution of $m$ with field and its saturation are possibly due to the reconstruction of FS by magnetic field. It may be mentioned that SdH and dHvA oscillation studies have probed the FS in the field and temperature region  where the value of $m$ is very close to 1. Further studies are necessary to resolve the issue of difference in Fermi surface geometry, constructed from zero-field probe and quantum oscillation technique, as mentioned in the introduction of the present manuscript and also to understand the role of magnetic field on charge scattering.\\
\subsection{Tuning  of charge carrier by In doping}
From the above discussion on field and temperature dependence of thermoelectric power, it is clear that one can tune the Fermi surface by applying magnetic field. As a result, $m$ increases from nearly zero to 1. As $S$ depends inversely on the carrier density, one expects that $S$ will decrease upon electron doping in Cd$_3$As$_2$. If the linear temperature dependence of $S$ persists with doping, it may be possible to tune the FS to make  $m$ negative. With this idea in mind, we have  doped a very small amount of In (2$\%$) at Cd site in Cd$_3$As$_2$.  To determine the carrier density, the Hall measurement was done at different temperatures in the range 2-300 K, as shown in Fig. 5(a). The density of electronic charge carrier is calculated to be $\sim$1.5$\times$10$^{19}$ cm$^{-3}$, which is higher than the typical carrier density in Cd$_3$As$_2$. At zero field, the resistivity ($\rho$$_{xx}$) of (Cd$_{0.98}$In$_{0.02}$)$_3$As$_{2}$ single crystal decreases monotonically with decreasing temperature down to 2 K  as shown in Fig. 5(b). The residual resistivity ratio [$\rho$$_{xx}$(300 K)/$\rho$$_{xx}$(2 K)] is about 7 at zero field. At 2 K, $\mu$ is calculated to be $\sim$6$\times$10$^{4}$ cm$^{2}$ V$^{-1}$ s$^{-1}$, which is comparable to that for the undoped crystal. Under application of magnetic field, resistivity shows a metal-semiconductor like crossover as in the case of parent compound \cite{tian}.  With the increase in field strength, this anomaly enhances and shifts towards higher temperature. Except very few  semimetals, like WTe$_{2}$, NbP, Bi$_{0.96}$Sb$_{0.04}$, etc. \cite{ali,shek,Yue}, conventional semimetals do not exhibit such behavior, which may be due to the opening of a gap at the Dirac point. Considering the thermally activated  type carrier transport as in the case of semiconductors, $\rho$$_{xx}$(T)=$\rho$$_{0}$exp(E$_{g}$/$\kappa_{B}$T), a very small energy gap was obtained from the slope of ln($\rho$$_{xx}$) vs $T^{-1}$ plots, as shown in Fig. 6. Inset shows the magnetic field variation of the induced energy gap. The gap increases rapidly with the increase in magnetic field.\\

\begin{figure}
\includegraphics[width=0.4\textwidth]{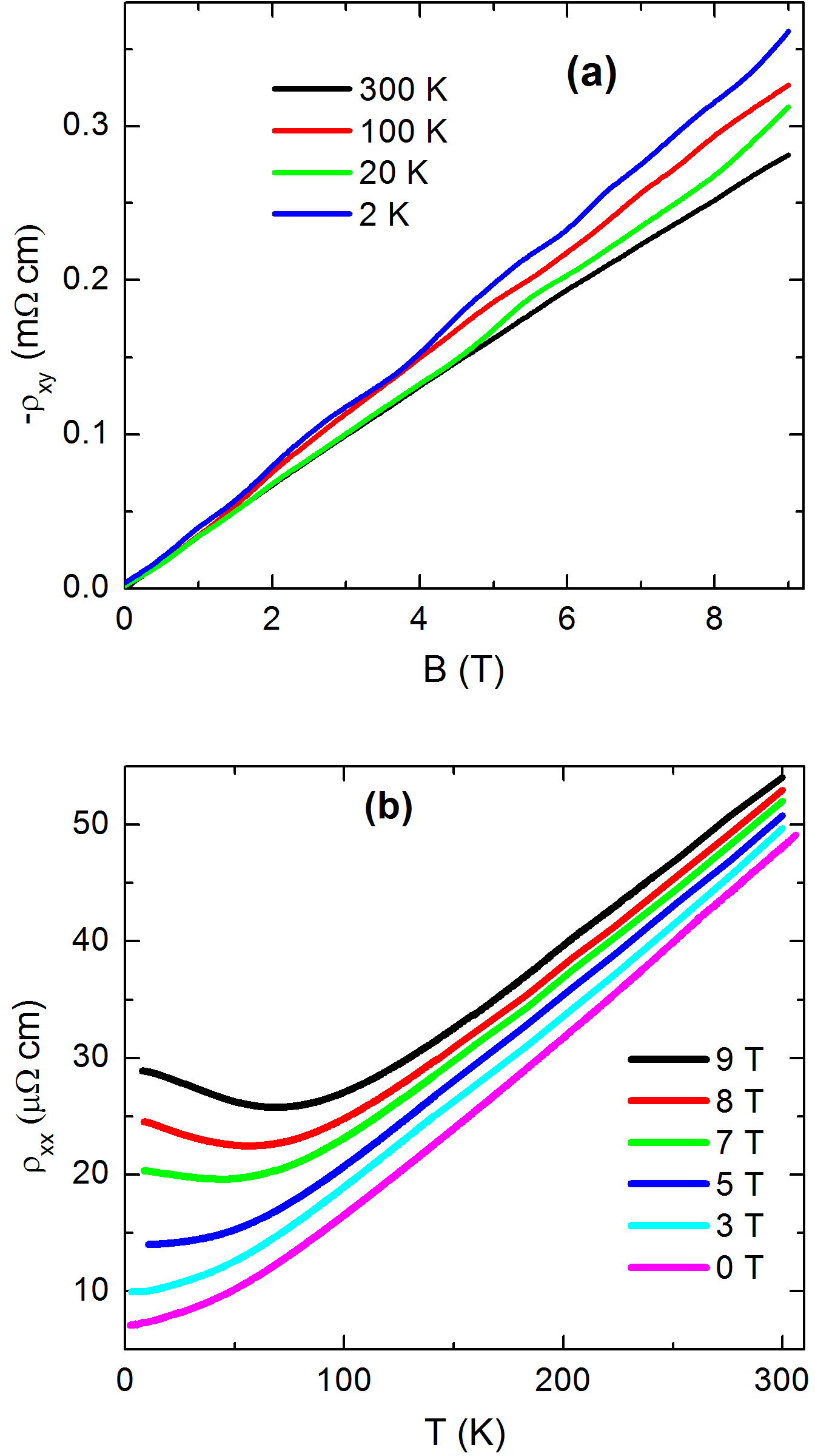}
\caption{(a) Field dependence of the Hall resistance ($R_{xy}$) at some representative temperatures and (b) temperature dependence of resistivity ($\rho$$_{xx}$) for different applied fields for (Cd$_{0.98}$In$_{0.02}$)$_3$As$_{2}$  single crystal.}\label{rh}
\end{figure}

\begin{figure}
\includegraphics[width=0.4\textwidth]{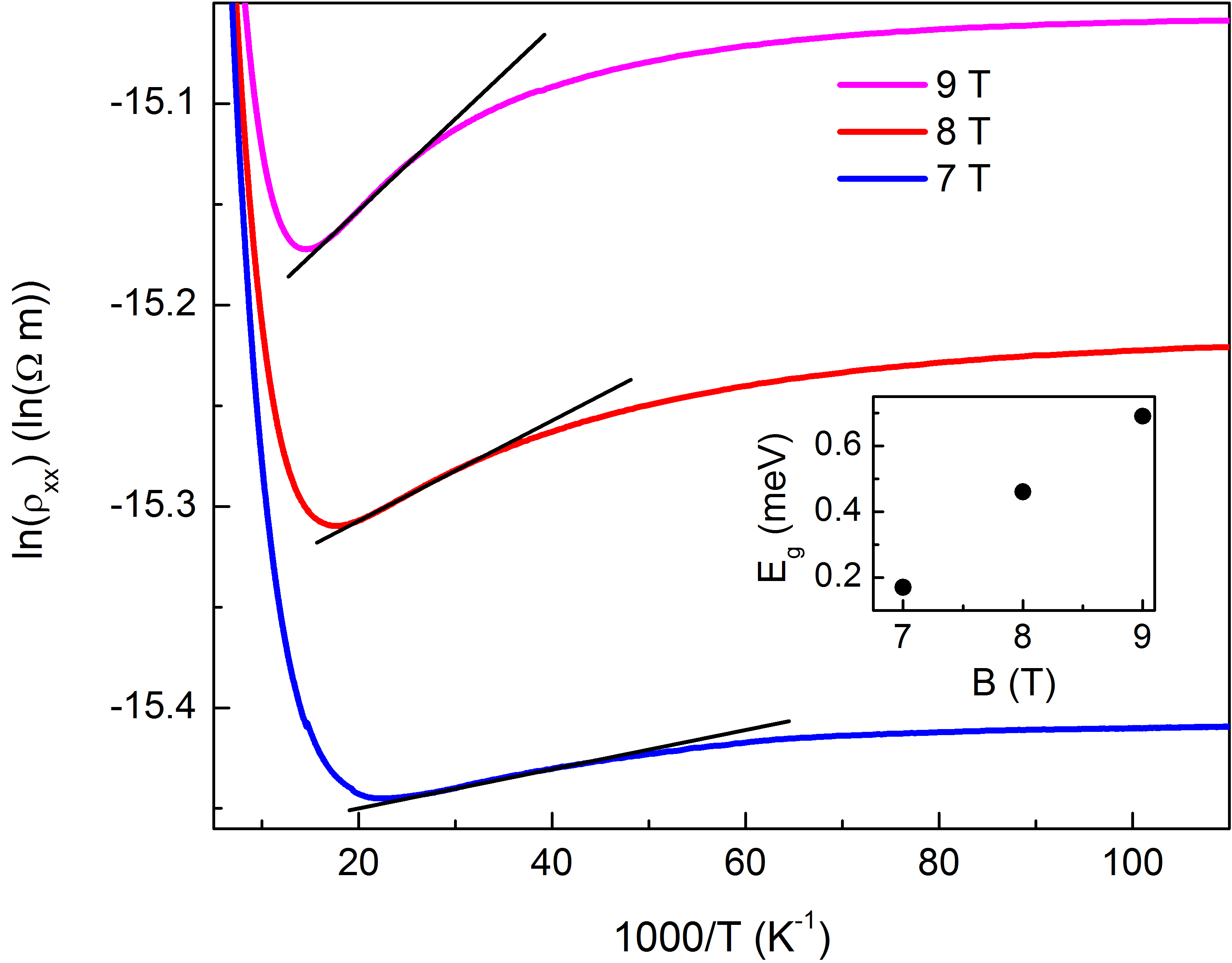}
\caption{(Color online) ln($\rho$$_{xx}$) versus 1000T$^{-1}$ plot for (Cd$_{0.98}$In$_{0.02}$)$_3$As$_{2}$ crystal. Using the slope in linear region, the thermal activation energy gap induced by the magnetic field has been calculated. Inset shows the field variation of the energy gap above 7 T.}\label{rh}
\end{figure}
\subsection{Analysis of magnetoresistance and Shubnikov-de Haas oscillations in (Cd$_{0.98}$In$_{0.02}$)$_3$As$_{2}$ crystal}
\begin{figure}
\includegraphics[width=0.4\textwidth]{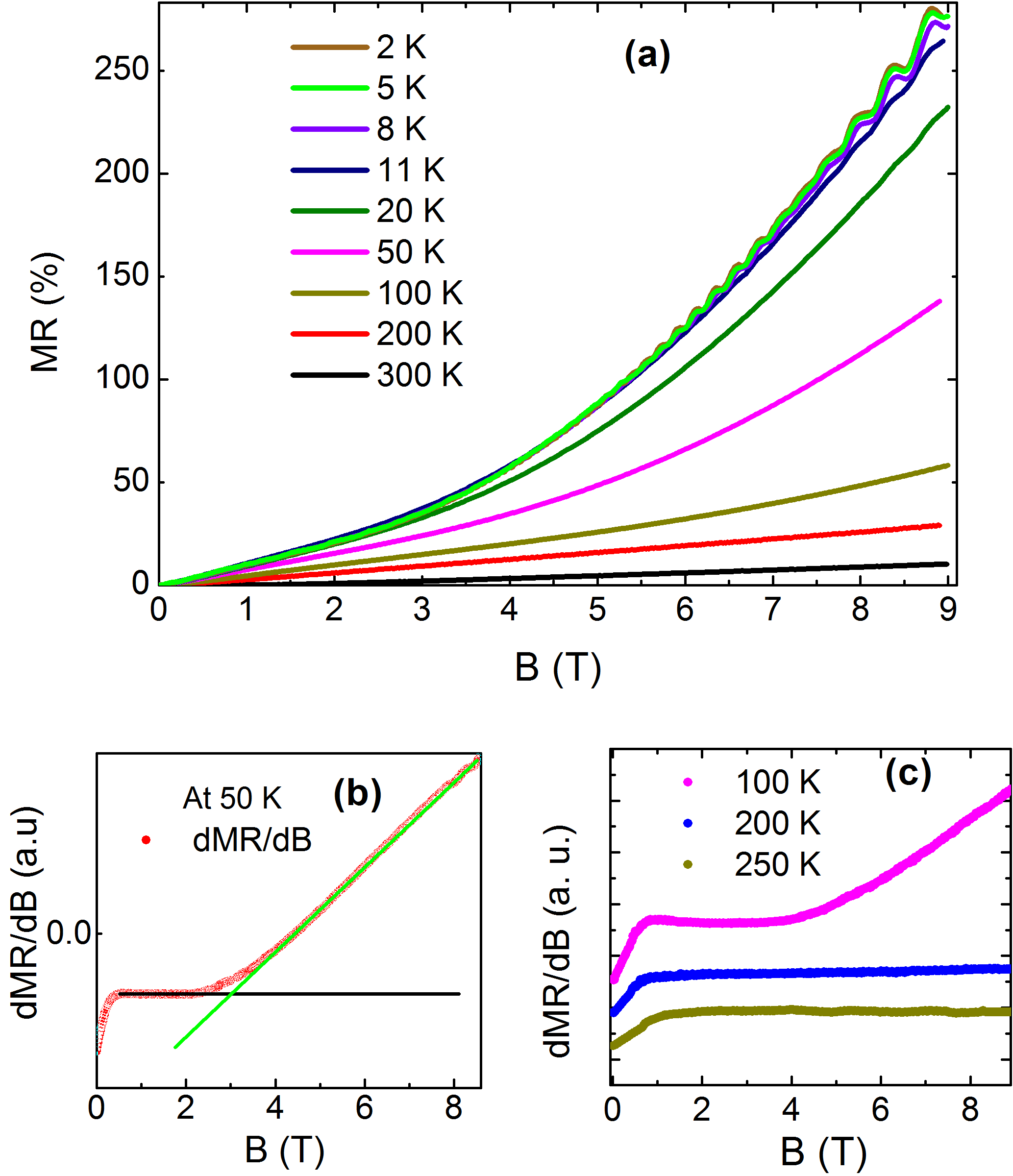}
\caption{(Color online) (a) Magnetoresistance  of (Cd$_{0.98}$In$_{0.02}$)$_3$As$_{2}$  single crystal at some selected temperatures between 2 to 300 K. (b) The first-order derivative of MR [d(MR)/dB] at 50 K. The black and green solid lines represents linear MR and quadratic MR region respectively. (c) d(MR)/dB at some representative temperatures above 50 K. }\label{rh}
\end{figure}
The transverse magnetoresistance  of (Cd$_{0.98}$In$_{0.02}$)$_3$As$_{2}$ single crystal is measured in the temperature range from 2  to 300 K with magnetic field along [100]  direction and current along [012] direction. The results are shown in Fig. 7(a). Unlike Cd$_{3}$As$_{2}$,  In-doped single crystal shows non-linear MR \cite{he,nara,Zhao,arnab}. At low temperature, MR is approximately linear only below a threshold value of magnetic field (except in a very narrow region around $B$=0), while it is quadratic at high field. The quadratic  component of MR gradually suppresses with increasing temperature. The first-order derivative of MR [d(MR)/d$B$] at 50 K has been plotted in  Fig. 7(b) to show the exact magnetic field dependence. The low-field region, which broadens with increasing temperature, is also present in Cd$_{3}$As$_{2}$ \cite{tian}. d(MR)/d$B$ is flat in the intermediate-field region and linear at high field, which correspond to linear and quadratic nature of MR, respectively. d(MR)/d$B$ vs $B$  in Fig. 7(c) shows that the high-field quadratic region gradually shrinks and the linear region gradually expands with increasing temperature. At 2 K and 9 T,  MR is about 280 \% which suppresses to only $\sim$10\% at 300 K.\\

In(Cd$_{0.98}$In$_{0.02}$)$_3$As$_{2}$ crystal, the Shubnikov-de Haas  oscillation has been observed up to 20 K. The oscillatory component of MR has been calculated by subtracting a smooth polynomial background from the total MR and is shown in Fig. 8. The amplitude of oscillation reduces with increasing temperature and  suppresses completely above 20 K. The fast Fourier transform of the oscillation, as shown in the inset of Fig. 8, reveals two distinct frequencies ($F$) at around 159.3  and 184.6 T, which indicate the presence of two Fermi surface cross sections perpendicular to the applied field direction [100]. The presence of two frequencies in Cd$_{3}$As$_{2}$ has been ascribed to the nesting of two equivalent ellipsoidal Fermi surfaces beyond the Lifshitz transition \cite{Zhao,arnab}. Employing the Onsager relation $F$$=$($\phi$$_0$/2$\pi$$^2$)$A_F$, the Fermi surface cross sections ($A_F$) perpendicular to [100] direction are calculated to be 1.52 $\times$ 10$^{-2}$ and 1.76 $\times$ 10$^{-2}$ ${\AA}^{-2}$ respectively, which are at least 3 times higher than the undoped compound \cite{he,nara,Zhao,arnab}. Assuming circular cross sections, the Fermi wave vectors ($k_F$) are determined to be $\sim$0.07 ${\AA}^{-1}$ and $\sim$0.075 ${\AA}^{-1}$, respectively.\\
\begin{figure}
\includegraphics[width=0.4\textwidth]{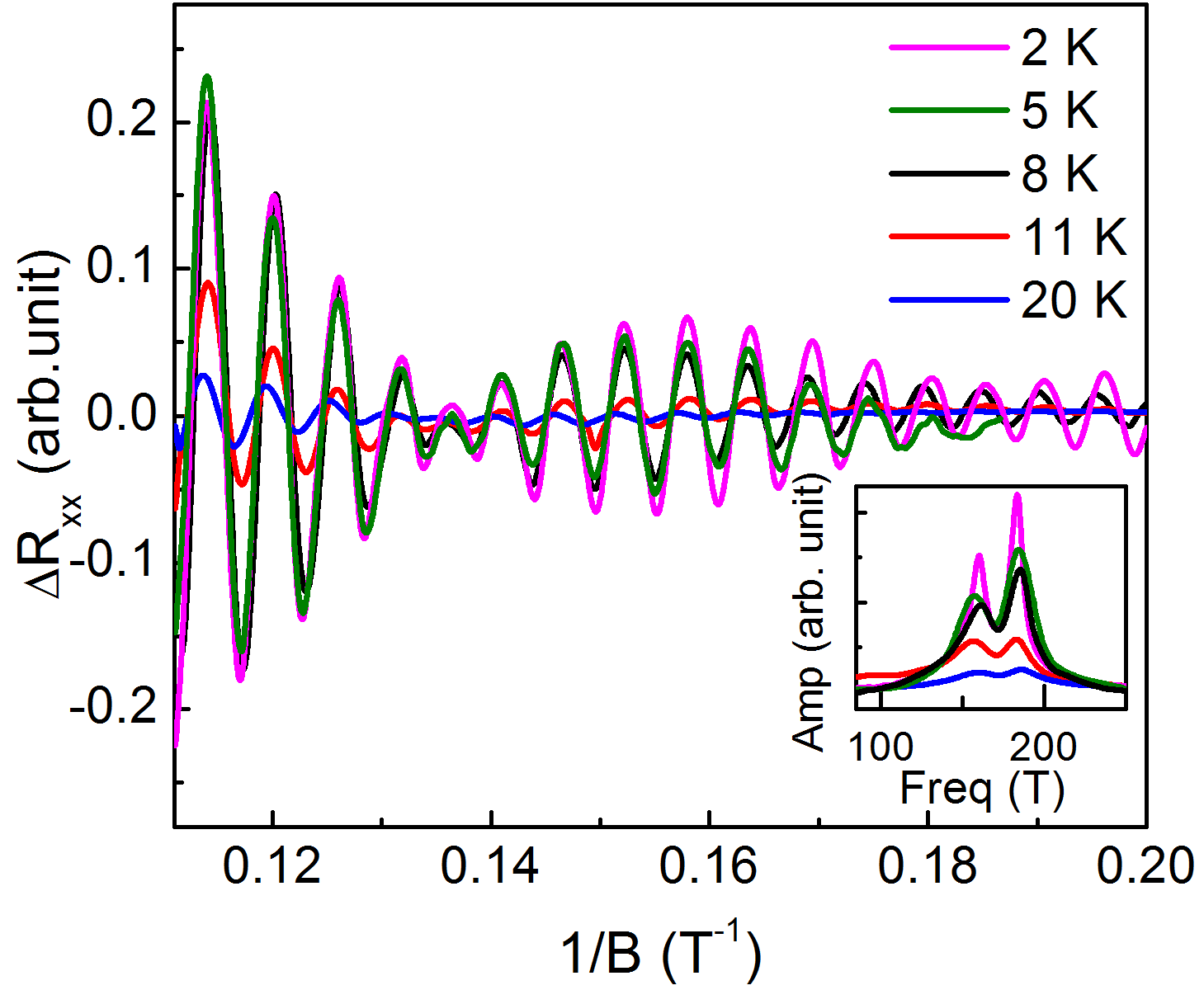}
\caption{(Color online) The oscillatory component $\triangle$R$_{xx}$  of MR  as a function of 1/$B$ at various temperatures with magnetic field along the [100] direction; the inset shows the oscillation frequency after fast Fourier transform in (Cd$_{0.98}$In$_{0.02}$)$_3$As$_{2}$. }\label{rh}
\end{figure}

The thermal damping of the  amplitude of oscillation $\Delta$R$_{T}$ [Fig. 9(a)] can be described by the temperature-dependent part of the Lifshitz-Kosevich formula:
\begin{equation}
 \Delta R_{T}=a\frac{2\pi^2k_BT/\hbar\omega_c}{\sinh(2\pi^2k_BT/\hbar\omega_c)},
\end{equation}
where $a$ is a  temperature-independent constant  and $\omega_c$ is the cyclotron frequency. The energy gap $\hbar\omega_c$ can be obtained by fitting the  amplitude of SdH oscillations with Eq. (4). The effective cyclotron mass of the charge carrier ($m^{\star}_{e}$) and  the Fermi velocity are obtained from the relations $\omega_{c}$$=$$eB/m^{\star}_{e}$ and $v_F$$=$$\hbar$$k_F/m^{\star}_{e}$ respectively. $m^{\star}_{e}$ is calculated to be $\sim$0.13 $m_e$, which is almost three times larger than that for the parent compound \cite{arnab}. $v_F$ is found to be $\sim$ 6.1$\times$ 10$^{5}$ m/s.\\

In Fig. 9(b), the oscillatory component of MR  is shown to be  fitted by the Lifshitz-Kosevich formula, $\Delta R_{xx}$=$A\exp(-cB)\cos2\pi(F/B+0.5+\beta)$, excluding the thermal damping term [Eq.(4)]. We have excluded the thermal damping term because it is a much slower varying function of $B$ than the other two terms. Here, $c$$=$2$\pi^{2}k_{B}T_Dm^{\star}_{e}/\hbar e$ and $2\pi\beta$ is the Berry's phase, where $\beta$ can take values from 0 to 1/2 (0 for the parabolic dispersion as in the case of conventional metals and 1/2 for the linear dispersion in 3D Dirac system).  Taking into account the presence of two frequencies, we have used superposition of two oscillatory components correspond to F$_{1}$= 159.3  and F$_{2}$=184.6 T to fit the experimental data. Initially, we have fixed $\beta$ to 0 for parabolic band. But, as shown in Fig. 9(b), the fitting to the experimental curve is much inferior. By tuning $\beta$ from 0 to 1/2 in successive steps, the fitting to the experimental data improves progressively. Figure 9(b) shows fitting with $\beta$=0.35 in the theoretical expression. With further increase in $\beta$ above 0.35, we observe that the fitting becomes inferior. $\beta$$>$0 demonstrates that the Fermi energy is in the linear dispersing region. This is fully consistent with the earlier STM \cite{jeon} and ARPES \cite{liu1} reports, which state that the linear dispersion in Cd$_{3}$As$_{2}$ persists up to as high as 500 meV from the Dirac point.\\
\begin{figure}
\includegraphics[width=0.4\textwidth]{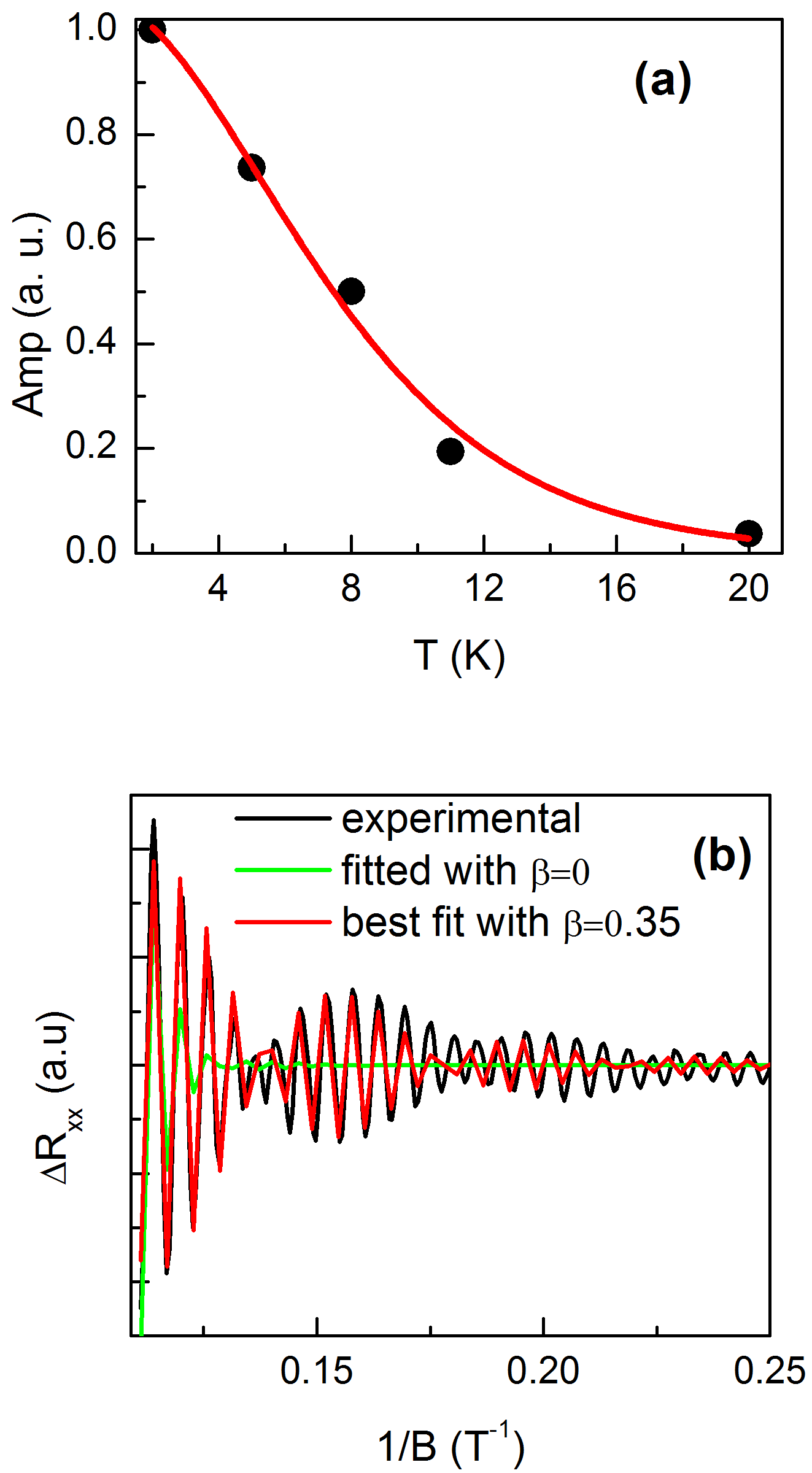}
\caption{(Color online) (a) Thermal damping of oscillation amplitude in (Cd$_{0.98}$In$_{0.02}$)$_3$As$_{2}$ crystal. Solid points are the experimental data and the red line is the fit to the experimental data. (b) The oscillatory component $\triangle$R$_{xx}$  of MR at 2 K (after subtracting a smooth background). Black, experimental curve; green, fit with $\beta$=0; red, fit with non-zero value of $\beta$.}\label{rh}
\end{figure}
\subsection{Temperature dependence of $S$ in (Cd$_{0.98}$In$_{0.02}$)$_3$As$_{2}$}
\begin{figure}
\includegraphics[width=0.4\textwidth]{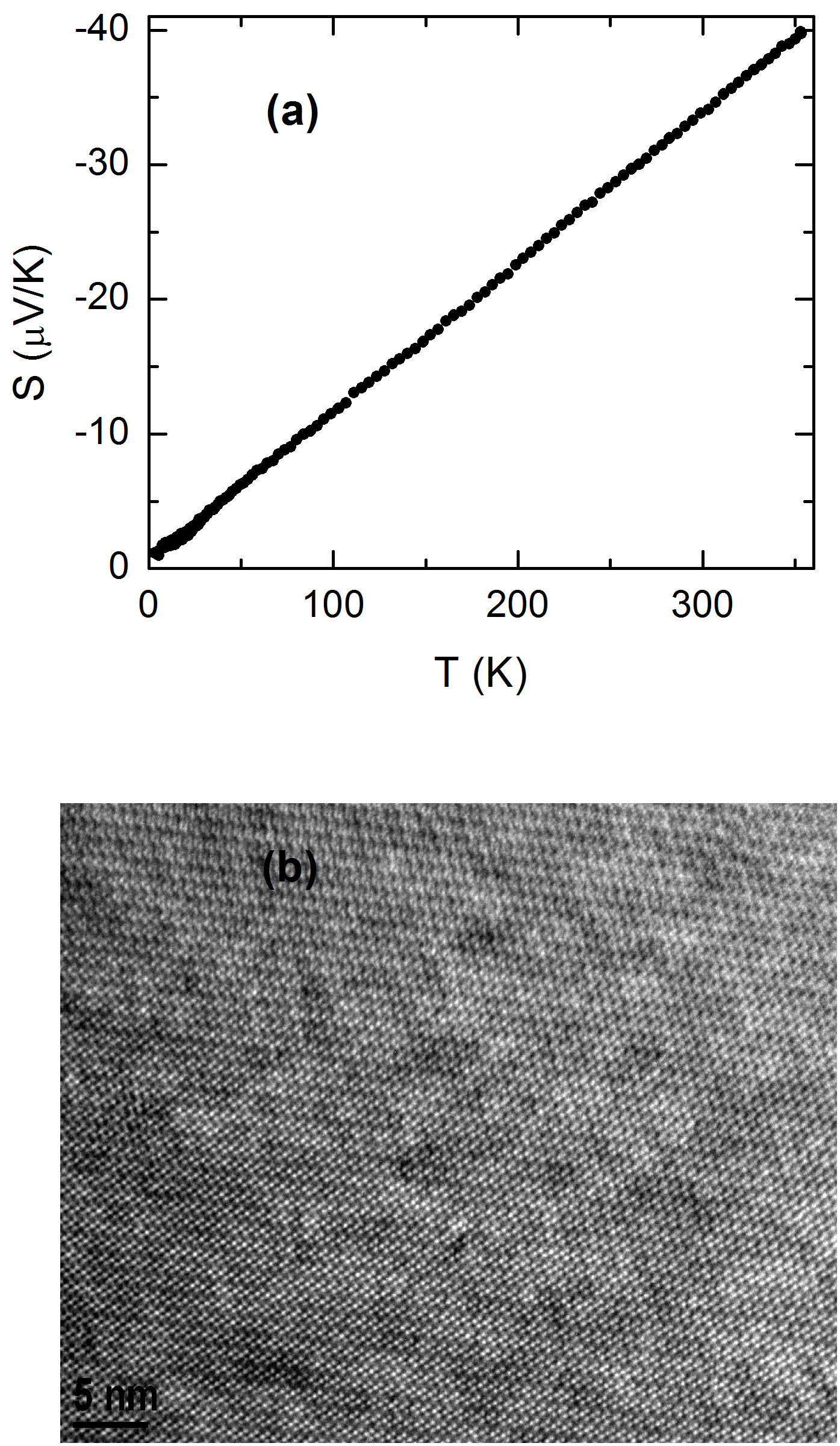}
\caption{(Color online) (a) Temperature dependence of the thermoelectric power ($S$) of (Cd$_{0.98}$In$_{0.02}$)$_3$As$_{2}$ up to 350 K. (b) High resolution transmission electron microscopy (HRTEM) image of the crystal.}\label{rh}
 \end{figure}
Fig. 10(a) shows the temperature dependence of the thermoelectric power of (Cd$_{0.98}$In$_{0.02}$)$_3$As$_{2}$ crystal. Similar to  undoped compound, $S$ exhibits linear $T$ dependence, whereas the value of $S$ reduces significantly upon carrier doping. Using the experimental values of $k_F$ and $m^{\star}_{e}$ in the Fermi energy expression for the relativistic excitation, $E_F$=$\hbar$$^{2}$$k_F^2/m^{\star}_{e}$, we get $E_F$$\sim$ 302 meV. Now, using the slope of $S$($T$) in Eq. (3), the scattering exponent $m$ is calculated to be $\sim$-0.6 for the relaxation process in (Cd$_{0.98}$In$_{0.02}$)$_3$As$_{2}$ crystal. For both acoustic phonon scattering and neutral white-noise short-range disorder, the value of $m$ is known to be -1 \cite{hwang2}. However, the high resolution transmission electron microscopy image of the crystal does not show any signature of short-range disorder [Fig. 10(b)]. Very clear selective area electron diffraction pattern obtained in HRTEM measurements also consistent with the absence of any kind of disorder [Fig. 1(a)]. So, the value of scattering exponent can be attributed to the emergence of acoustic phonon scattering.\\
\subsection{Magnetotransport and thermoelectric properties of  (Cd$_{0.96}$In$_{0.04}$)$_3$As$_{2}$ crystal}
With the intension to tune the scattering exponent further, we have doped 4\% In at the Cd site in Cd$_{3}$As$_{2}$. Fig. 11(a) shows  the temperature dependence of resistivity for (Cd$_{0.96}$In$_{0.04}$)$_3$As$_{2}$ crystal. From the figure, it is clear that $\rho$$_{xx}$($T$), in absence of magnetic field, shows metallic behavior (d$\rho$$_{xx}$/d$T$$>$0) down to lowest temperature similar to 2\% In-doped crystal. However, the value of $\rho$$_{xx}$ at a given temperature is larger than that for $x$=0.02 sample but smaller than $x$=0 sample. The value of residual resistivity ratio ($\sim$5) is smaller to that for both $x$=0 and $x$=0.02 samples. So, the electrical resistivity and residual resistivity ratio exhibit non-monotonic dependence on In content. We have also measured the Hall resistivity to determine the carrier concentration in this sample. The Hall resistivity has been plotted with $B$ in Fig. 12(a) at two representative temperatures, 2 and 300 K. Figure shows that $\rho_{xy}$ is independent of temperature and $\rho_{xy}$ vs $B$ is linear up to 9 T. From the slope of $\rho_{xy}$ vs $B$ plot, the carrier density is calculated to be $\sim$2.4$\times$10$^{19}$ cm$^{-3}$. The deduced value of $n$ is larger than the carrier density for $x$=0 and 0.02, implying In doping continue to increase carrier concentration. However, the carrier mobility  reduces to $\sim$1.1$\times$10$^{4}$ cm$^{2}$ V$^{-1}$ s$^{-1}$. From the above discussion, it is evident that disorder dominates the charge conduction mechanism in Cd$_{3}$As$_{2}$ above a certain In doping level, in spite of increase in carrier density.\\

In order to understand the effect of magnetic field on resistivity, we have also measured $\rho$$_{xx}$ at different applied  fields [Fig. 11(a)]. Figure 11(a) shows  that the resistivity enhances with the application of  magnetic field as in the case of $x$=0 and 0.02 samples. Similar to $x$=0 and 0.02, $\rho$$_{xx}$ for $x$=0.04 also shows the field-induced metal-semiconductor like crossover with decreasing temperature.  However, this phenomenon is weaker than that observed in 0 and 0.02 samples. MR as a function of  magnetic field has also been measured and plotted in Fig. 12(b). However, for this sample, we have not  observed any SdH oscillation up to 9 T magnetic field and temperature down to 2 K.  We have already shown  that the effective mass of the carrier increases significantly with the increase in carrier density resulting from In doping.  Also, the conductivity for the 4\% In-doped sample reduces due to the increase of disorder. The absence of quantum oscillation within the measured magnetic field range is possibly due to the increase in carrier effective mass and disorder with In doping. But, surprisingly, Fig. 12(b) shows large and nonsaturating MR for (Cd$_{0.96}$In$_{0.04}$)$_3$As$_{2}$ like the undoped sample. At 9 T, the MR is as high as $\sim$ 1650\% at 2 K and $\sim$ 250\% at room temperature. However, the nature of MR is weakly superlinear compared to weakly sublinear MR in undoped Cd$_{3}$As$_{2}$. As pointed out by Parish and Littlewood \cite{parish} through statistical simulation, large spatial fluctuations in mobility due to presence of disorder, can lead to large linear magnetoresistance. This benefit of imperfection was first experimentally realised in doped sample of Ag$_{2}$Se and Ag$_{2}$Te by several groups \cite{husmann,hu}. Recently, fluctuation in electron mobility due to collision with randomly distributed low-mobility islands (i.e., disorders), has been ascribed as a possible source of large and linear magnetoresistance in Cd$_{3}$As$_{2}$ \cite{nara}. It has also shown that the value of MR scales with the mobility of charge carrier. As shown in Table I, with further indium doping in Cd$_{3}$As$_{2}$, the large linear MR is recovered and the mobility of charge carrier is reduced from what is observed in 2\% In-doped sample.  This disorder-induced MR in 4\% In-doped sample is consistent with the above mentioned statistical model.  Controlling MR by tuning disorder through doping may be a potential route for the construction of magnetic field sensors.\\

\begin{figure}
\includegraphics[width=0.4\textwidth]{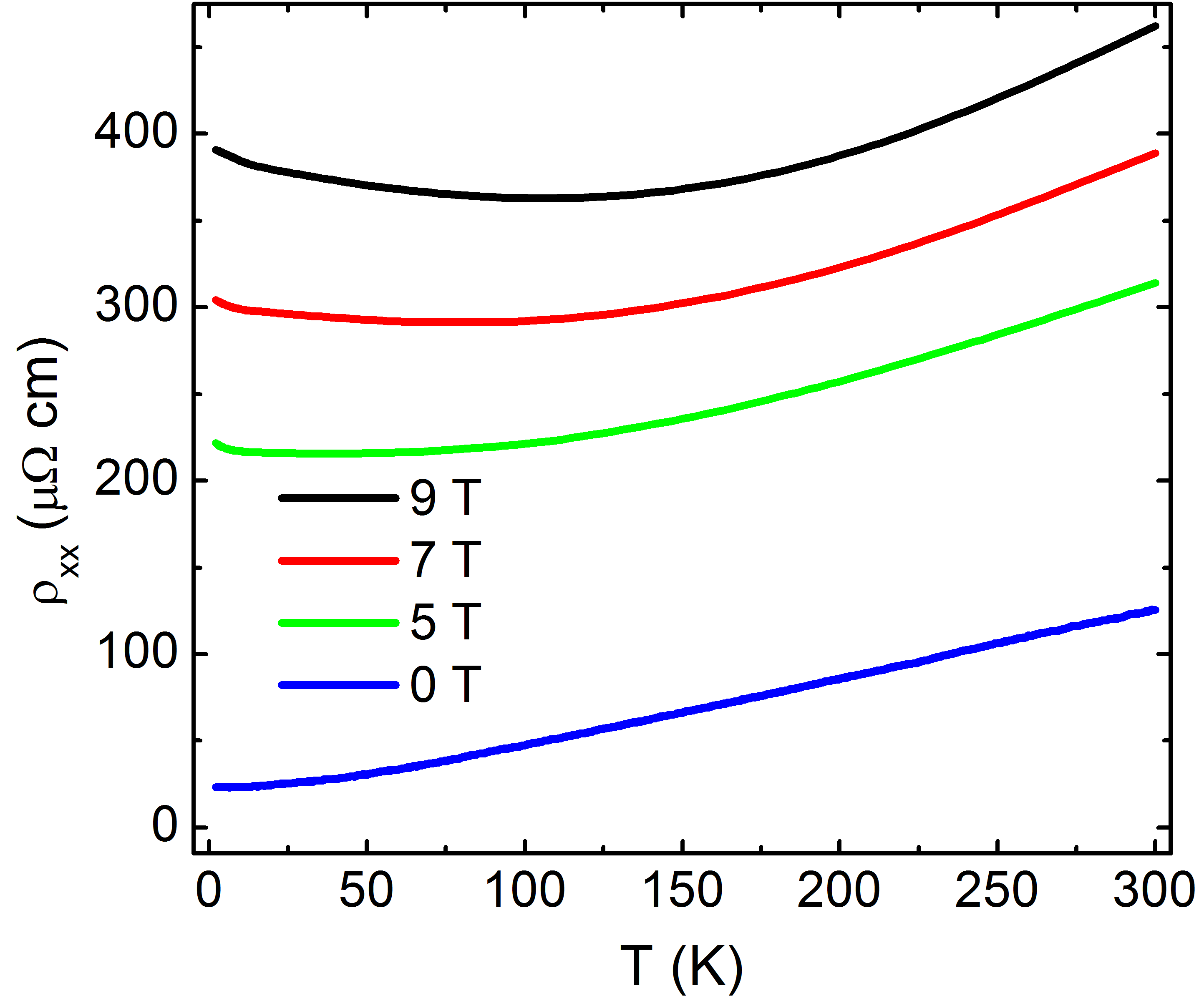}
\caption{Resistivity ($\rho$$_{xx}$) as a function of temperature for (Cd$_{0.96}$In$_{0.04}$)$_3$As$_{2}$.}\label{rh}
\end{figure}

\begin{figure}
\includegraphics[width=0.4\textwidth]{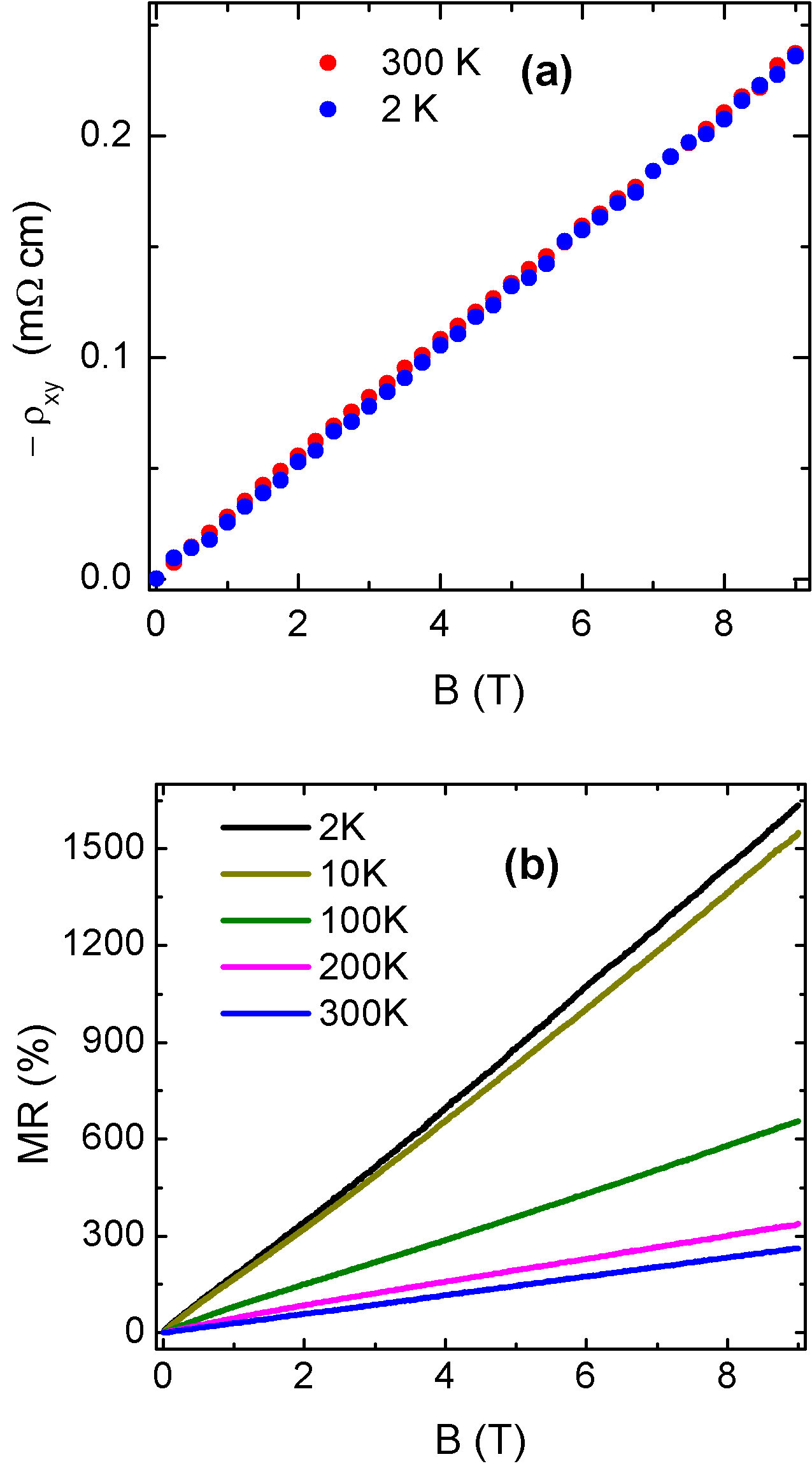}
\caption{(a) Field dependence of the Hall resistance ($R_{xy}$) at 2 and 300 K. (b) Magnetoresistance of (Cd$_{0.96}$In$_{0.04}$)$_3$As$_{2}$ at a few representative temperatures from 2 to 300 K.}\label{rh}
\end{figure}

\begin{figure}
\includegraphics[width=0.4\textwidth]{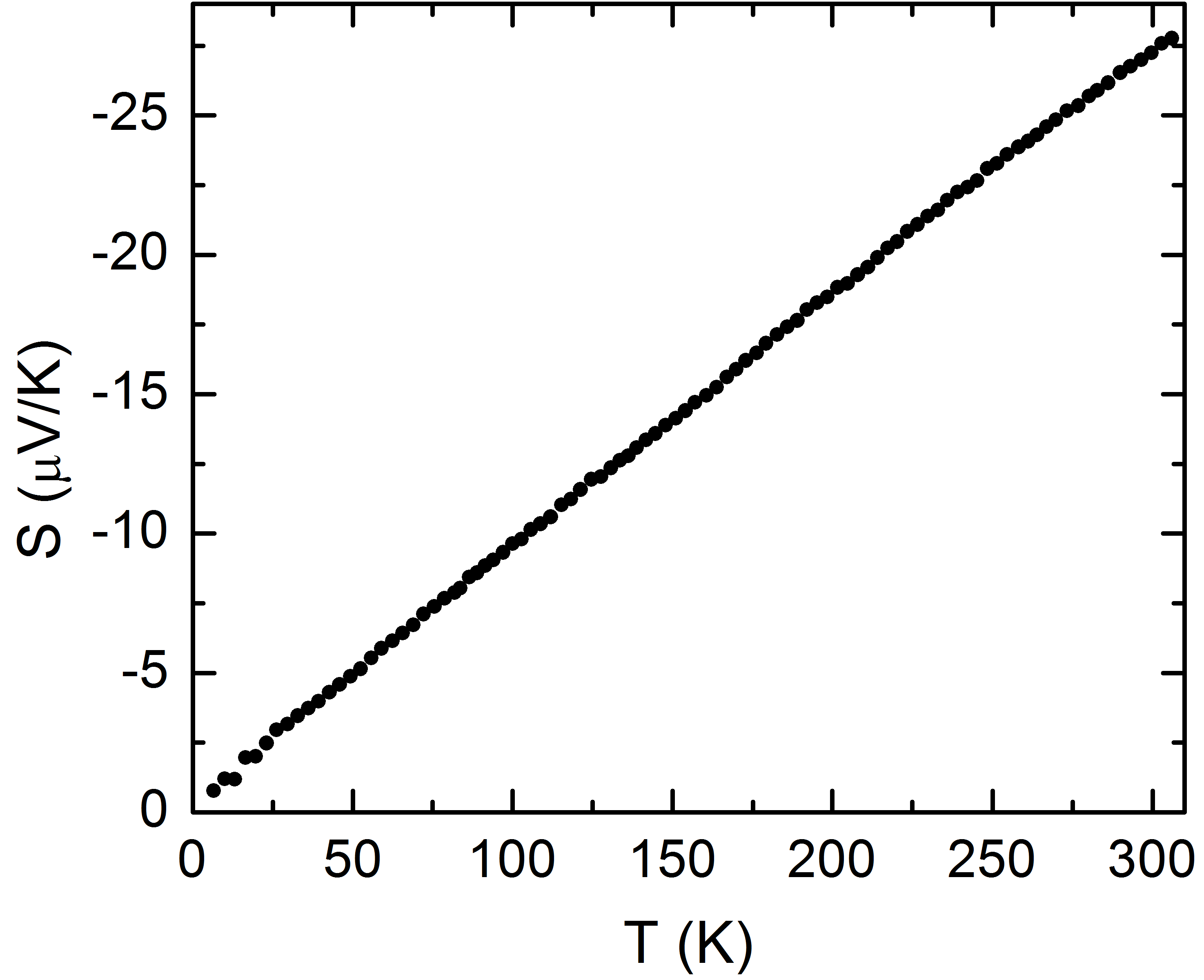}
\caption{(Color online) Temperature dependence of the thermoelectric power ($S$) of (Cd$_{0.96}$In$_{0.04}$)$_3$As$_{2}$ up to 300 K.}\label{rh}
\end{figure}
In Fig. 13, the Seebeck coefficient for (Cd$_{0.96}$In$_{0.04}$)$_3$As$_{2}$ is plotted  as a function of temperature up to 300 K. Consistent with the increase in carrier density, the value of $S$ reduces further. Similar to undoped and 2\% In-doped sample, $S$ vs $T$ is linear throughout the temperature range, i.e., the Mott semiclassical formula [Eq. (2)] is obeyed. However, due to the absence of quantum oscillations in the magnetoresistance data, it is not possible to comment whether 0.04 In-doped sample hosts Dirac semimetallic phase or not. Also, we cannot deduce $m$ using Eq. (3) due to the lack of knowledge on $E_F$. \\

\textbf{TABLE I:} The values of Seebeck coefficient at 300 K ($S_{300K}$), resistivity at 2 K ($\rho_{2K}$), carrier density ($n$), carrier mobility ($\mu$) and magnetoresistance at 2 K and 9 T for Cd$_{3}$As$_{2}$, 2\% indium-doped [Cd$_{3}$As$_{2}$(I)] and 4\% indium-doped [Cd$_{3}$As$_{2}$(II)] samples.
\begin{center}
\begin{tabular}{||c c c c c c||}

\hline
  & $S_{300K}$ & $\rho_{2K}$ & n & $\mu$ & MR\\
  & $\mu$V/K & $\mu$$\Omega$-cm & 10$^{18}$cm$^{-3}$ &  10$^{4}$cm$^{2}$/Vs &  \%\\
\hline
Cd$_{3}$As$_{2}$ & 60 & 70 & 6.8 & 1.3 & 1600\\

Cd$_{3}$As$_{2}$(I) & 34 & 7 & 15 & 6 & 280\\

Cd$_{3}$As$_{2}$(II) & 27 & 23 & 24 & 1.1 & 1650\\
\hline

\end{tabular}
\end{center}

\section{CONCLUSION}
In conclusion, the Seebeck coefficient shows linear temperature dependence over a wide range, which is in agreement with the  Mott's relation.
The  signature of three-dimensional linear dispersion in Cd$_3$As$_2$ has been clearly reflected from the magnetic field dependence of $S$. The relaxation process of charge carrier in Cd$_{3}$As$_{2}$ is found to be extremely sensitive to magnetic field and carrier doping. Analysis reveals that the scattering time evolves from being nearly energy independent to becoming linearly dependent on energy as the magnetic field increases.  Fermi surface is strongly affected and the scattering time enters into the inverse energy-dependent regime with 2\% indium doping at Cd ion position. Further doping (4\%) increases disorder in the system and SdH oscillation is no more traceable down to 2 K and up to 9 T applied magnetic field. Surprisingly, this higher doped sample shows large and linear magnetoresistance like the undoped compound. The observed behaviour of MR is consistent with the statistical model, which states that large spatial fluctuation in carrier mobility due to presence of disorder, can generate large linear MR. This disorder controlled MR in Cd$_3$As$_2$ may be useful for the fabrication of magnetic field sensor device.\\
\section{ACKNOWLEDGMENTS}
We thank Arindam Midya, Susmita Roy, and Arun Paul for their help during measurements.

\end{document}